\documentclass[10pt, aps, prl, superscriptaddress, nofootinbib, 
               amsmath, amssymb, twocolumn,
               preprintnumbers, showpacs,
               raggedbottom,
               floatfix]{revtex4-1}

\usepackage[T1]{fontenc}
\usepackage{textcomp}
\usepackage{etoolbox}
\usepackage{etextools}
\usepackage{comment}
\AtBeginDocument{\graphicspath{{/home/rishi/work/ambar/figures/}}}

\usepackage[acronym]{glossaries}
\glsdisablehyper
\makeglossaries
\newacronym{ope}{OPE}{one pion exchange}
\usepackage[utf8]{inputenc}

\usepackage[breaklinks]{hyperref}
\usepackage{graphicx}
\usepackage{dcolumn}
\usepackage{calc}
\usepackage{pdfcomment}
\usepackage{color}

\newcommand{\GeV}{{\rm{GeV}}}

\begin{document}

\title{Unravelling Medium Effects in Heavy Ion Collisions with Zeal}

\author{Rajiv Gavai}
\affiliation{TIFR, Homi Bhabha Road, Navy Nagar, Mumbai 400005, India}
\author{Ambar Jain}
\affiliation{Indian Institute of Science Education and Research Bhopal,
Bhopal Bypass Road, 
Bhopal 462066, India}
\author{Rishi Sharma}
\affiliation{TIFR, Homi Bhabha Road, Navy Nagar, Mumbai 400005, India}
\date{\today}

\begin{abstract}
\noindent
We propose a new observable, called zeal, to analyze events with jets in
heavy ion collisions.   The observable measures how a thermal medium affects
the multiplicity and distribution of energetic particles in a jet.  Using few
known models for energy loss and jet quenching, we demonstrate its capability
to distinguish the physics of these models.
\end{abstract}
\preprint{TIFR/TH/15-26}

\maketitle
\glsresetall

Since the seminal proposal~\cite{bjorken1982energy} of using energy loss of fast
particles and the related jet-quenching as a probe to study the nature of the
hot medium formed in heavy ion collisions, a lot of experimental and theoretical
studies have enriched our understanding of the medium produced in relativistic
heavy ion collisions~\cite{Stasto:2007ir,Vitev:2014kha}. For instance, $R_{AA}$,
obtained from the ratios of single particle inclusive transverse momentum
($p_T$)-spectra of nucleus-nucleus ($AA$) to that of suitably normalized
proton-proton ($pp$) spectra shows a large suppression at
RHIC~\cite{STAR:2004,PHENIX:2006,STAR:2007} and at the LHC~\cite{ATLAS:2013}.
Arguing it to stem from the leading particle in the corresponding jets, this has
been identified as the shining example of jet quenching by the medium,
especially since no such suppression in seen in ratios constructed for $pA$
collisions.

Several models of suppression for leading partons of a jet in a deconfined thermal medium
have been proposed to account for these data. The essential picture is that the
momentum of the leading parton transverse to its motion is broadened due to
kicks from the medium. The broadening per unit distance
($\hat{q}=\frac{k_T^2}{\lambda}$) is related to the energy loss and is often
used to characterize the interaction of leading partons with the thermal medium. 

The momentum broadening in the medium at a given temperature depends on the
detailed description of the medium. Several
groups~\cite{Gyulassy:2001,Salgado:2003gb,Qin:2009gw,Liu:2006ug,jeon:2005}
have calculated the suppression of leading partons in QGP . These involve
different models for the medium and the calculation of the energy loss.
(For a comparison between the approaches see Refs.~\cite{Armesto:2011ht}.)
Remarkably, a successful description of data seems feasible using models 
which treat the medium as a weakly coupled quark-gluon plasma (QGP), such as the
GLV~\cite{Gyulassy:2001} formalism or the AMY formalism~\cite{jeon:2005} as well
as models based on the assumption that the medium is strongly coupled (sQGP),
which treat $\hat{q}$~\cite{Salgado:2003gb,Wang:2001ifa,Qin:2009gw} as a
non-perturbative parameter, or evaluate it using techniques suitable for
strongly coupled theories~\cite{Liu:2006ug}. It is clearly desirable to
find ways to distinguish between them.  The extended $p_T$-range at LHC is
already a welcome help in that direction.  One may also look for new observables
where these models, having already been constrained by $R_{AA}$, differ in 
their predictions for them.

Energetic particles in quantum chromodynamics (QCD) come not isolated but as a
part of a shower of several collimated particles known as jets. This is true
already in the absence of the medium because
of~\cite{Gribov:1972,altarelli:1977,Dokshitzer:1977} splitting of partons
before fragmentation.  In the QGP, interactions with the medium enhance these
processes. 

LHC~\cite{ATLAS:2013,Aad:2014wha,CMS:2014,ATLAS:2015} has ushered us in a new
era of quantitative analyses of reconstructed jets, inviting thus a direct
scrutiny of the original jet quenching ideas since one deals with truly fast
particles.  A jet is defined by clustering of observed particles in an event
which depends on a parameter that is correlated with the angular width of jets
or equivalently the cone radius $R=\sqrt{\Delta\eta^2+\Delta\phi^2}$.  The
corresponding ratio $R^{\rm jets}_{AA}$ depends, however, on the cone
size \cite{ATLAS:2013}.  Intuitively this seems
understandable~\cite{Vitev:2008rz,He:2011pd} since jets with larger $R$ capture
more momentum flow. However, it obscures a comparison with theory unlike
$R_{AA}$ due to a $R$-dependent variation in $p_T$-suppression. Furthermore,
one has to contend with a background subtraction in heavy ion collisions,
adding an extra source of difference in a comparison with $pp$ collisions (or
central vs peripheral collisions employed commonly).  

In this Letter, we propose a new observable for studying jets with the aim of
minimizing, if not eliminating, the background subtraction and the cone size
$R$-dependence and yet capturing properties distinguishing the structure of 
jets in $AA$ and $pp$ collisions. We define {\it{transverse zeal}}\footnote{Here we use only the transverse momenta, hence the adjective
``transverse''. Another definition, involving net magnitude of
momenta, will give a different distribution. This may be useful to
study the jet substructure involving energetic particles in $pp$
collisions. Here on, we simply refer to $Z$ as the zeal.} of a jet as follows,
\begin{equation}
\begin{split}
Z = -[\log(\sum_i  \exp{[-p_T/(\hat{n}_T\cdot \vec{p}^i)]})]^{-1}~\label{eq:zeal}\;,
\end{split}
\end{equation}
where, the sum runs over the particles in the identified jet, $p_T$ is the
magnitude of the total transverse momentum of the jet, $\hat{n}_T$ is the
transverse direction of the jet, and $\vec{p}^i$ is the momentum of the $i$th
particle in the jet. In the extreme case of only a single leading particle in
the jet, $Z =1$, whereas in the other extreme case  of a jet of particles with
typical energies of the order of the temperature $T$, $Z \ll 1$.  
 
Since one expects that  jets in $AA$ collisions will typically consist of a
larger number of particles sharing the net $p_T$ due to broadening caused by
both (a) medium induced bremsstrahlung and (b) thermal kicks in the medium, they
will have a smaller zeal compared to the $pp$ case. The difference may not be as
stark as the extreme example above due to possible hadronization
effects.  Nonetheless we expect that the zeal distribution observed in heavy-ion
collisions to be shifted towards lower values compared to that in $pp$ (or $pA$)
collisions. Moreover, the shift will depend on the spectrum of particles
populating the jets in heavy-ion collisions compared to the spectrum in $pp$ (or
$pA$) collisions and will therefore encode quantitative information about how
the medium affects the propagation of an energetic particle. As one needs a
cone size $R$ only to determine the $p_T$ of the jet, and as the background
will be dominated by particles with much smaller transverse momenta, we expect
the zeal distribution to be much less sensitive to both these factors.

One may estimate the $p_T$ of the jet by using typical jet identification
algorithms for the $AA$ environment~\cite{Cacciari:2010te}.  Alternatively, one can use the near side
jet to identify the transverse momentum in the hard process and then analyze the
zeal distribution of the away side jet.~\footnote{One appealing possibility is to
use $\gamma$ or $Z$ tagged jets, subject to the size of the initial state
radiation~\cite{Neufeld:2010fj} for the away side particle.} For a
large enough cone radius $R$, the value of the jet momentum $p_{\rm{jet}}$
should not be particularly sensitive to $R$.  The typical cone radius for $pp$
collisions beyond which $p_{\rm{jet}}$ saturates is about $0.6$.  $R$ may be
larger in $AA$ collisions because of (a) momentum broadening transverse to the
jet $p_T$ due to the medium (b) greater probability of gluon emission. 
For a large enough $R$, though, the sensitivity of the extracted value of $p_T$
should be reduced.

Since the zeal weighs the contribution of multiple energetic particles in the
jet, it is more sensitive to the details of the distribution of the momentum
between the constituent particles compared to the $R_{AA}$ which only follows
the leading particle. Thus it certainly has a complimentary, perhaps greater, 
capability in differentiating models.  For example, if a
large fraction of the energy of the leading particle in an away side jet is lost
to many thermal partons, the typical zeal values will be nearly zero. On the
other hand if a few energetic partons carry away the energy then the typical
zeal will still be finite. These scenarios may be relevant for the sQGP and wQGP
respectively, subject to hadronization effects which themselves will decrease the
zeal.  

\begin{figure}[htbp]
\includegraphics[width=0.40\textwidth,clip=]
{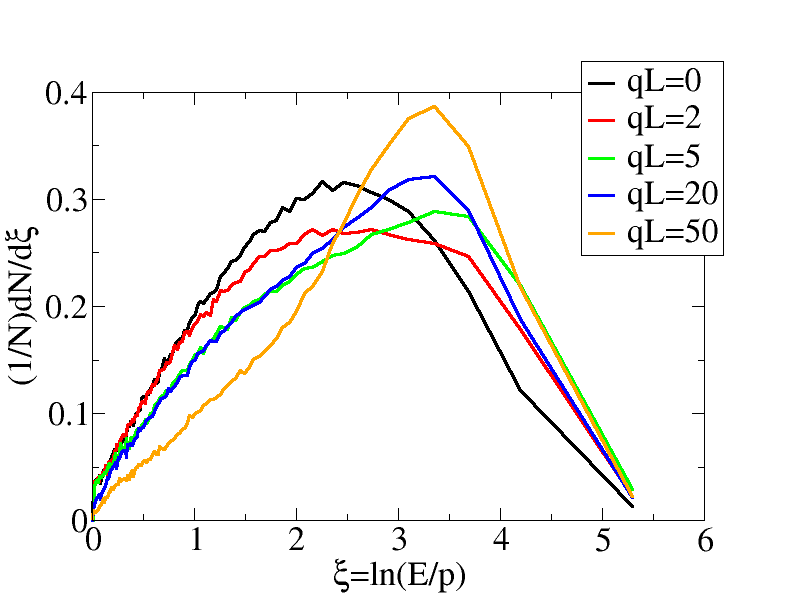}
  \caption{Distributions of partons in $100$GeV jets for various 
$\hat{q}$ as a function of $\xi$. The corresponding distributions for 
$10$GeV jets are similar but narrower and peak at smaller $\xi$. See 
Ref.~\cite{Armesto:2009} for details.\label{fig:partons}}
\end{figure}

We test the viability of these ideas by investigating the difference between
zeal distributions of $pp$ and $AA$ collisions for a few popular models of
jet-quenching in the literature in different kinematic regimes.  Our goal here
is to elucidate the main features of the zeal distribution in heavy ion collisions
and to check if it is a useful observable to characterize jets in the heavy-ion
environment, and we find that it is. Computation of zeal of Eq.~(\ref{eq:zeal})
needs the spectrum of particles which constitute a jet.  While it can be taken
from the experimental data or theoretical model simulations, we employ the
published momentum distribution of particles in jets predicted by a few models.

Let us first consider a model discussed in Ref.~\cite{Armesto:2009}. The paper
describes implementation of a medium modified QCD-splitting function in a Monte
Carlo parton cascade for an energetic parton travelling through the fireball.
The modification of the splitting depends on the jet-quenching parameter 
$\hat{q}$, which is related to transverse broadening in the thermal medium.  
We chose it as a useful model since the spectrum of partons in a shower in both
$pp$ and $AA$ collisions are published, allowing us to evaluate the zeal in
the two cases and compare. We will also later consider some more popular
models which are: GLV (based on
~\cite{Gyulassy:1999zd,Gyulassy:2001,Gyulassy:2001nm}), ASW (based on
~\cite{Arnold:2001ms,Arnold:2002ja}), AMY (based
on~\cite{Wiedemann:2000za,Salgado:2003gb,Armesto:2003jh}). 

\begin{figure*}[htbp]
\includegraphics[width=0.40\textwidth,clip=]
{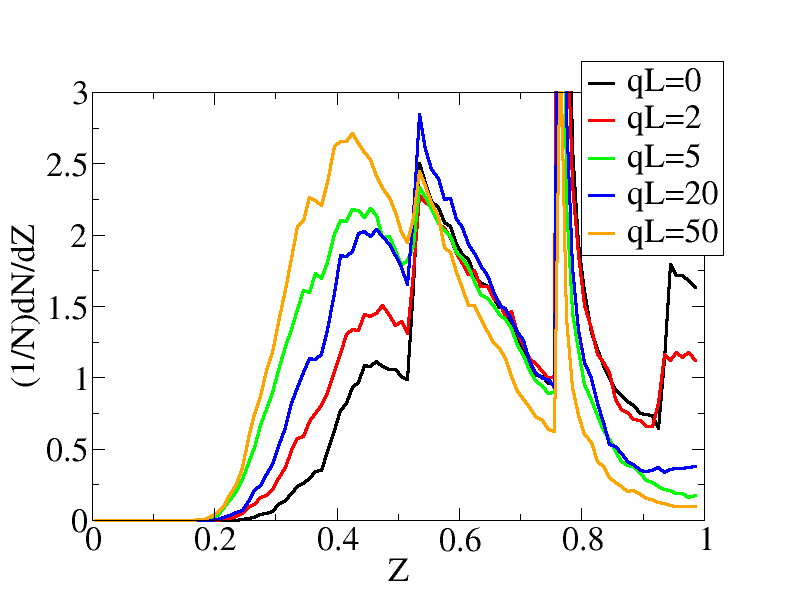}
\includegraphics[width=0.40\textwidth,clip=]
{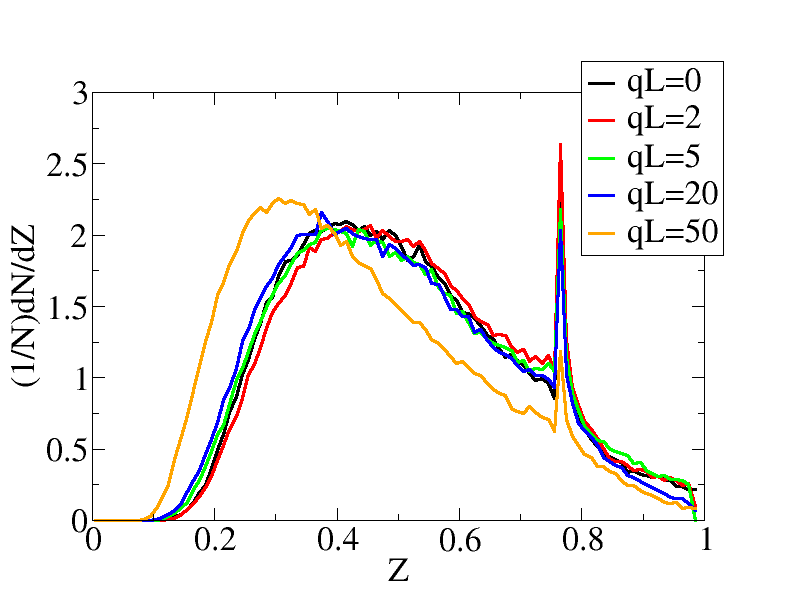}
  \caption{Zeal distributions for $10$GeV and $100$GeV jets for various 
  $\hat{q}$.\label{fig:zeals}}
\end{figure*}

One can obtain the detailed spectrum of particles in a jet in this model by
simulating events using the Monte Carlo QHERWIG and identifying jets, as in
~\cite{Armesto:2009}. For simplicity, here we generate an ensemble of jets as
follows.  Consider a parton shower with total energy $E$. In QHERWIG, this is
created by injecting a gluon with energy $E$ into the medium. We partition this
energy into partons carrying a fraction $x=\exp(\xi)$ chosen randomly according
to a probability distribution function which is identical to the predicted
distribution function of partons given in Ref.~\cite{Armesto:2009} (see Fig. $4$
in Ref.~\cite{Armesto:2009}).  This ensures that to a good approximation the
average (over a large number of jets constructed in this manner) distribution of
partons as a function of the fractional energy carried is the same as the
predicted distribution using QHERWIG. 

These distributions are given in Ref.~\cite{Armesto:2009} for $E=10$GeV and
$E=100$GeV and for $\hat{q}L=0$, $\hat{q}L=2\GeV^2$, $\hat{q}L=5\GeV^2$,
$\hat{q}L=20\GeV^2$, $\hat{q}L=50\GeV^2$, where $L$ is the path length of
the parton in the medium. Using the software ``g3data'' we extracted the data
from the plots and employ as the probability density function for the procedure
described above. Our resultant distributions of partons in a jet of energy
$100$GeV, averaged over an ensemble of $200,000$ parton showers is given in
Fig.~\ref{fig:partons}. A comparison with Fig. $4$ in Ref.~\cite{Armesto:2009}
shows that the obtained distribution is similar to the predicted distribution in
QHERWIG. The same is also true for the distributions for $10$GeV jet energy
which we have not shown here. 

If we consider events of equal
centrality (or equivalently system size) one can study the range of medium-jet
coupling from weak to the strong coupling. Alternately, if we know the value of
$\hat{q}$, the range of simulations covers central to peripheral events
(assuming that the effect of the expected reduction in $\hat{q}$ due to a
slight lowering of the temperature for more peripheral collisions is not
substantial and the dominant effect is the change of geometry).

\begin{figure}[htbp]
\includegraphics[width=0.40\textwidth,clip=]
{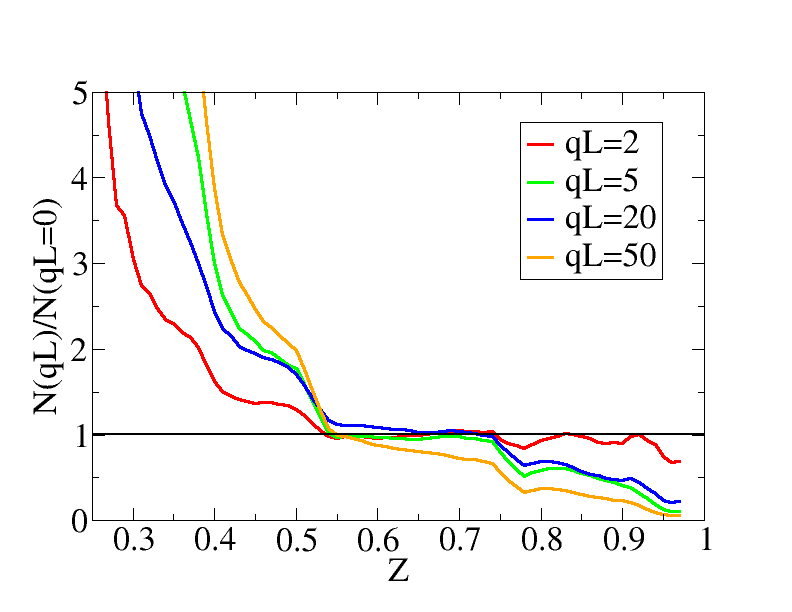}
\includegraphics[width=0.40\textwidth,clip=]
{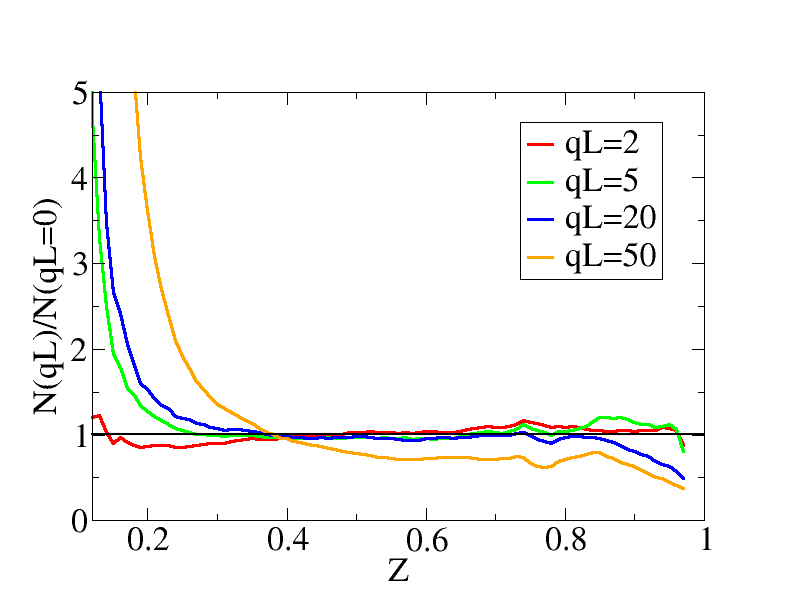}
  \caption{Ratios of distributions for $10$GeV (upper panel) and $100$GeV
  (lower panel) jets with the distribution for ${\hat{q}}L=0$ distribution for
  various $\hat{q}$.~\label{fig:zealratios}}
\end{figure}

Fig.~\ref{fig:zeals} displays  zeal-histograms of events for a single gluon jet
in the medium with increasing $\hat{q}$ for both $10$ and $100$GeV. 
The case with $\hat{q}=0$ corresponds to the $pp$ collision where the
distribution for $\xi$ is sharply peaked at $0.765$. This is associated with
jets where two particles share the bulk of the energy of the jet. Likewise,
there is also a three-particle multiplicity peak clearly visible in the data at
about $Z\sim 0.54$ in particular for $10$GeV. We expect these few particle
structures to be smoothened out in detailed simulations when hadronization is
taken into account. 

We note that in Fig.~\ref{fig:zeals} smaller zeal values get populated for both
$10$GeV and $100$GeV jets, as the medium broadening effect increases. In addition,
the height of the two-multiplicity peak at $Z=0.765$ decreases.  The difference between
$\hat{q}L=0$, $\hat{q}L=20$ and $\hat{q}L=50$ scenarios is easily noticeable.
The broad peak shifts towards smaller zeal values becoming somewhat narrower at
larger $\hat{q}L$ values and the two-multiplicity peak vanishes.  We
expect the main feature that the zeal distribution moves towards smaller
values with larger $\hat{q}L$ to be robust and worth exploring further.

In order to contrast the $\hat{q}L=50$ and $\hat{q}L=0$ cases better we construct
their ratios, as shown in Fig.~\ref{fig:zealratios}. These may be recognised to
be similar to the $R_{CP}$, where one studies centrality dependence of the
medium effects, or $R_{AA}$ which are main tools to study jet-quenching
experimentally since systematic effects cancel out in them.  As for $R_{CP}$, no
change in zeal distribution due to medium effects will correspond to a constant
line at unity for the ratio.  We observe a net decrease in events with zeal
values above $Z\simeq 0.4$ ($Z\simeq0.5$) for $E=100$GeV ($E=10$GeV) and net
increase in zeals below, with increasing values of $\hat{q}L$. For a fixed
$\hat{q}$, this amounts to increasing the path length or more central
collisions.  We also note that the ratio decreases as the zeal
increases. In particular, very interestingly, the ratio of the few particle
structures fall with increasing $\hat{q}L$ in a manner similar to the smooth
region of the plot.  

Different models of medium effects will have different predictions for the zeal
yields. Comparison with experimental data for the zeal distribution can possibly
rule some out or verify them.  It is well known (for eg.  \cite{Armesto:2011ht})
that the value of $\hat{q}$ may not be a good way to characterize different
models because the same value of $\hat{q}$ gives rise to different medium
induced bremsstrahlung rates in different formalisms, and consequently different
values of $R_{AA}$.  It may be better to use the zeal distributions
for a fixed centrality to differentiate various theoretical models of
jet quenching.  The medium induced bremsstrahlung spectra are directly related
to those of relatively energetic particles moving along the jet
direction, at least before hadronization. Therefore,  it may turn out better to
compare the zeal distributions predicted for various models
for the respective medium effects tuned  to obtain the correct $R_{AA}$ and see
whether they can further assist in distinguishing the models by a comparison
with the experimental results. 

\begin{figure}[htbp]
\includegraphics[width=0.40\textwidth,clip=]
{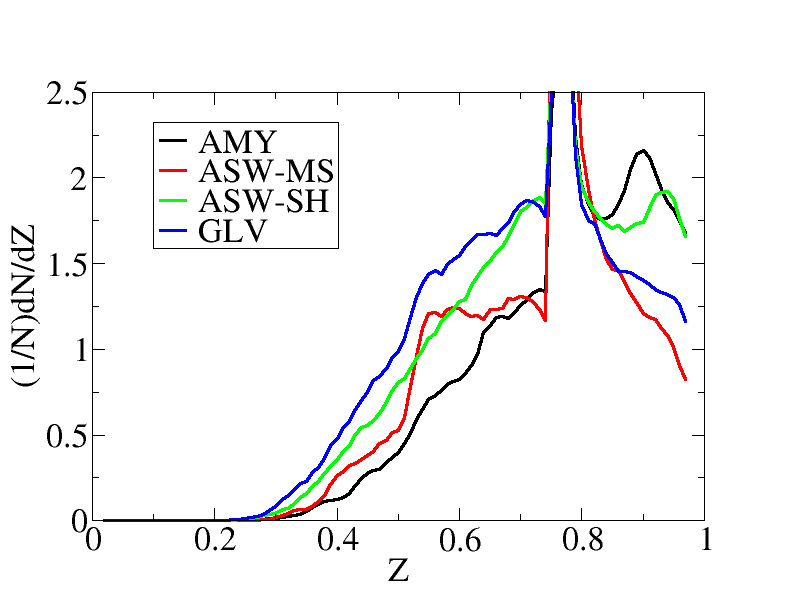}
\includegraphics[width=0.40\textwidth,clip=]
{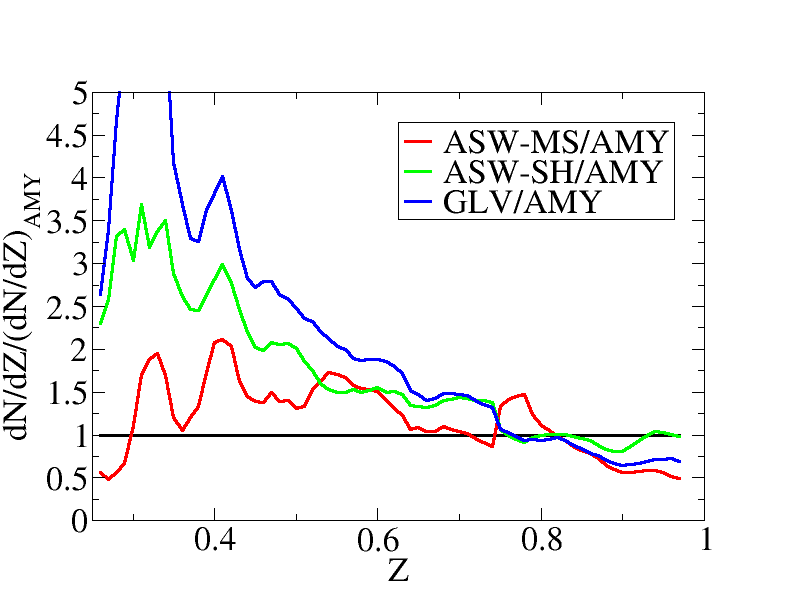}
  \caption{The upper panel shows zeal distributions for $20$GeV for various
  models described in Ref.~\cite{Armesto:2011ht}. The lower panel shows ratios
  of the other models with AMY. ~\label{fig:zealmodels}}
\end{figure}

Fig. $19$ of Ref.~\cite{Armesto:2011ht} provided us the necessary input for our
procedure above.  It compares the models
\cite{Gyulassy:1999zd,Gyulassy:2001,Gyulassy:2001nm,Arnold:2001ms,Arnold:2002ja,Wiedemann:2000za,Salgado:2003gb,Armesto:2003jh}
by tuning the model parameters such that the value of $R_7$ --- the suppression
of a $p_T$ spectrum falling with the power $1/p_T^7$ --- is $0.25$ when the
leading parton traverses a QGP brick of length $L=5$fm.  We generate ensembles
of events for the initial injected energy of $20$GeV partitioned into
partons carrying a fraction $x$ chosen according to the distribution given in
Ref.~\cite{Armesto:2011ht}.  Fig.~\ref{fig:zealmodels} exhibits our results for
the zeal distribution for different models obtained by following the same
procedure as described above.  We find it very satisfying that even when the
models are tuned to give the same suppression, the zeal distribution is able to
distinguish between them, in particular showing a significant difference between
the AMY and the GLV models. 


In summary, we propose a new jet observable, the zeal, defined by
Eq.~(\ref{eq:zeal}) to unravel the medium effects.  It depends on the transverse
momentum distribution of particles in the jets, and is therefore more
discriminating than $R_{AA}$ of the leading partons.   Its 
advantage is that it weighs the energetic partons more heavily and hence is
particularly sensitive to the processes that lead to the energy loss of the
leading partons. For frequent medium induced bremsstrahlung with several gluons
carrying a tiny fraction of the energy of the leading partons, the peak of the
zeal distribution should move towards lower zeal values  unlike the case where
induced bremsstrahlung is rare and the emitted gluons carry significant
fractions of the energy of the leading partons. 

Its second advantage is likely to be its smaller sensitivity to the
background and the cone radius $R$.  Since one can use large values of
jet cone radii $R$ to extract the $p_T$ of the jet to define zeal, it may be
further useful since jets are expected to be wider in $AA$ compared to $pp$
collisions due to broadening.  This will reduce the systematics associated with
the extraction of the $p_T$ of the jets, and hence the calculation of zeal.  It
is also infrared safe (an extra soft parton $i$ will not change the zeal
significantly because its contribution goes as $\exp(-p_T/p_T^i)$) and is not
affected significantly by accidental inclusion of some particles which are part
of the thermal medium. 

Our results suggest that the zeal distribution is sensitive to medium effects,
shifting it to smaller zeal values with increasing centrality. We also see that
it can be used to distinguish models of jet quenching.  It will be interesting
to study how these results are affected by hadronization.  We are currently
pursuing this study and its application to the experimental data.

We acknowledge the kind hospitality of WHEPP-13 where this work was initiated.
RVG and AJ acknowledge gratefully the support of the J. C. Bose Fellowship and
Ramanujam Fellowship respectively.

\bibliographystyle{apsrev4-1}
\bibliography{local}

\end{document}